\documentclass[aps,prb,twocolumn]{revtex4}

\usepackage{graphicx}

\begin{document}


\title{Heat transport through a Josephson junction}

\author{Dmitry Golubev$^1$, Timoth\'e Faivre$^2$, and Jukka P. Pekola$^2$}
\address{
$^1$ Karlsruhe Institute of Technology (KIT), Institute of Nanotechnology, 76021 Karlsruhe, Germany \\
$^2$ Low Temperature Laboratory (OVLL), Aalto University School of Science, P.O. Box 13500, 00076 AALTO, Finland
}

\begin{abstract}
We discuss heat transport through a Josephson tunnel junction under various bias conditions. 
We first derive the formula for the cooling power of the junction valid for arbitrary time dependence of the Josephson phase.
Combining it with the classical equation of motion for the phase, we find the time averaged cooling power
as a function of bias current or bias voltage. We also find the noise of the heat
current and, more generally, the full counting statistics of the heat transport through the junction.
We separately consider the metastable superconducting branch of the current-voltage characteristics
allowing quantum fluctuations of the phase in this case. This regime is experimentally attractive since the junction 
has low power dissipation, low impedance and therefore may be used as a sensitive detector. 
\end{abstract}

\maketitle

\section{Introduction}

Thermal effects in metallic nanostructures
are subject of intense  experimental and theoretical research\cite{review}.
In this paper we consider a particular example of such a structure -- a Josephson
junction between two superconductors. Although Josephson junctions are very important building
blocks of many low temperature devices, their heat transport properties
are still not fully understood and there exist only a few theoretical studies of this issue  in the literature.
Guttman {\it et al}\cite{Guttman}
have derived the heat current through a Josephson junction with
different temperatures of the leads and biased below  critical current so that the Josephson phase was fixed.
Subsequently, Zhao {\it et al}\cite{Zhao1,Zhao2} have corrected a sign error in their result  
and extended it to the case of arbitrary transparency of the barrier
between the superconducting leads. Their theory have been confirmed in a nice recent experiment
by Giazotto and Martinez-Perez \cite{GM}.

Frank and Krech \cite{FK} have considered a voltage biased Josephson
junction and demonstrated that the superconducting lead with a smaller gap gets cooled in a
certain range of bias voltages. They have derived an expression for the
cooling power of the junction, which was a straightforward generalization of the well known
result for the normal metal - insulator - superconductor (NIS) tunnel junction \cite{Nahum,Leivo}.

In this paper we develop a general approach to this problem, which covers the two regimes discussed above
as limiting cases. First, we allow the Josephson phase
to depend on time in an arbitrary way and derive the expression for the heat current, or cooling power, under these conditions. 
Next, we determine the actual time dependence of the phase by
solving the differential equation of motion, which describes phase dynamics under given bias conditions
and in the presence of noise of the environment\cite{Likharev, Barone}. More specifically,
we use the model of resistively and capacitively shunted Josephson junction (RCSJ model) with the shunt resistance
much smaller than both the resistance quantum and the resistance of the junction itself (see Fig. \ref{fig_schematics}). 
Afterwards, we find the cooling power averaged over time and noise
which can be measured in an experiment.

At bias currents below the critical one we use a more elaborate technique treating
the Josephson phase as a quantum operator. It is necessary to ensure the detailed balance
between forward and backward rates of the quasiparticle tunneling. In this limit we study the dependence
of the cooling power on the bias current and on the  difference between the temperatures of the environment
and of the superconducting leads. We believe that the low bias regime is experimentally most interesting
because the junction in this case has low noise, low impedance and low thermal conductance, which opens
up an opportunity of its application as a sensitive detector.

Apart from that, we study the noise of the heat current. It is also an important
characteristic of the junction in view of its potential detector applications.
    
On the technical side, we perform the usual second order perturbation theory in tunnel Hamiltonian
connecting the two superconducting leads. In doing so 
we use the full counting statistics (FCS) approach to the heat transport which has been recently
successfully applied to similar problems\cite{Laakso1,Laakso}.
The method has an advantage of providing the information about the cooling power of the junction,
its noise and, eventually, all cumulants of the heat current at one step.     
   
The paper is organized as follows: in Sec. II we introduce the model; in Sec. III
we derive FCS of the junction; in Sec. IV we derive the general expression for the cooling power; 
in Sec V we consider the heat current noise; in Sec. VI we study average cooling power under
various bias conditions; and, finally, in Sec. VII we summarize our results.
Some details of the calculations are given in the appendices.

\section{Model}

\begin{figure}
\includegraphics[width=8cm]{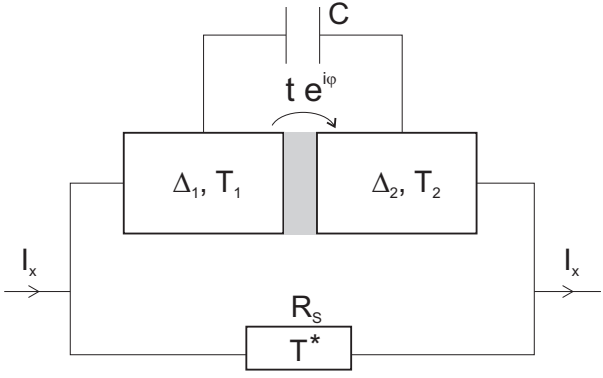}
\caption{A Josephson junction between two superconductors with the gaps $\Delta_1, \Delta_2$ and temperatures $T_1,T_2$.
The junction is shunted by a capacitor $C$, a resistor $R_S$ and the whole circuit is biased by a current $I_x$.
The dynamics of this circuit is described by RCSJ model (\ref{RSJ}). }
\label{fig_schematics}
\end{figure}

We consider a Josephson tunnel junction with the normal state resistance $R$ connecting two superconducting leads with 
the energy gaps $\Delta_1,\Delta_2$ and the temperatures $T_1,T_2$.
The junction is shunted by a capacitor $C$ and by a resistor $R_S$ kept at temperature $T^*$, as  shown in Fig. \ref{fig_schematics}. 
The ohmic resistor provides the simplest model for the electromagnetic environment of the junction. 
Our primary goal is to find the heat current through the junction or, more precisely,  
the cooling power of the superconductor 1 $P^{(1)}$. 
The latter is defined as the amount of energy extracted from the superconductor 1 per unit time.
We will also be interested in the noise of the cooling power. 
 
The junction is described by the Hamiltonian
\begin{eqnarray}
H=H_{1} + H_{2} + H_T,
\end{eqnarray}
where
\begin{eqnarray}
H_{1} = \sum_{k}\bigg[\sum_{\sigma=\uparrow,\downarrow}\epsilon_{1k} c^\dagger_{\sigma k} c_{\sigma k}  
+\Delta_1 c_{\uparrow k}c_{\downarrow -k} + \Delta_1 c_{\uparrow -k}^\dagger c_{\downarrow k}^\dagger \bigg]
\end{eqnarray}
is the Hamiltonian of the superconductor 1,
\begin{eqnarray}
H_{2}=\sum_{k}\bigg[\sum_{\sigma=\uparrow,\downarrow}\epsilon_{2k} a^\dagger_{\sigma k} a_{\sigma k}  
+\Delta_2 a_{\uparrow k}a_{\downarrow -k} + \Delta_2 a_{\uparrow -k}^\dagger a_{\downarrow k}^\dagger \bigg]
\end{eqnarray}
is that of the superconductor 2, and
\begin{eqnarray}
H_T=\sum_{kn}\sum_{\sigma=\uparrow,\downarrow} 
\big[ t_{kn} e^{i\hat\varphi/2} a^\dagger_{\sigma n} c_{\sigma k} + t_{kn}^* e^{-i\hat\varphi/2}c^\dagger_{\sigma k} a_{\sigma n}  \big],
\end{eqnarray}
is the tunnel Hamiltonian. Note that by means of a suitable gauge transformation we 
made the superconducting order parameters in the leads, $\Delta_1,\Delta_2$, real.  
After this transformation the Josephson phase $\hat\varphi$ is moved into the phase of the tunneling amplitude.
We will treat the phase $\hat\varphi$ as a quantum operator, which is indicated by the hat.
We further assume that the quasiparticle distribution functions in the leads are Fermi functions,
\begin{eqnarray}
f_j(E)=1/\left(1+e^{E/k_BT_j}\right),\;\; j=1,2.
\label{fj}
\end{eqnarray}
Thus we ignore possible charge imbalance\cite{Tinkham} in the leads.


Next, we will assume that the shunt resistance is much smaller than the  resistance quantum $R_k=h/e^2$ and the resistance
of the junction
\begin{eqnarray}
R_S\ll R_k, R.
\label{condition_RS}
\end{eqnarray}
In this case, and at sufficiently high bias current and/or temperature, one can
treat the Josephson phase as a classical variable. Its dynamics is then
described by the RCSJ differential equation\cite{Likharev,Barone}
\begin{eqnarray}
C\frac{\hbar \ddot\varphi}{2e}+\frac{1}{R_S}\frac{\hbar \dot\varphi}{2e}+I_C\sin\varphi = I_x +\xi_S,
\label{RSJ}
\end{eqnarray} 
where $\xi_S$ is the noise of the resistor $R_S$. 
At high temperature it is white Gaussian noise with the correlator
\begin{eqnarray}
\langle \xi_S(t)\xi_S(t')\rangle = \frac{2k_BT^*}{R_S}\delta(t-t').
\label{FDT}
\end{eqnarray}
The critical current of the junction, $I_C$, is  determined by the integral\cite{AB}
\begin{eqnarray}
I_C  =
\frac{1}{e R} \int_{\Delta_1}^{\Delta_2} dE \frac{\Delta_1\Delta_2\big[1-2f_1(E)\big]}{\sqrt{E^2-\Delta_1^2}\sqrt{\Delta_2^2-E^2}}.
\label{Ic} 
\end{eqnarray}

Here we have chosen $\Delta_1<\Delta_2$, i.e. superconductor 1 is supposed to be weaker than superconductor 2.

Equation (\ref{RSJ}) describes the motion of an effective particle with the coordinate $\varphi$
in the tilted Josephson potential
\begin{eqnarray}
U(\varphi)=-E_J\cos\varphi -{\hbar I_x \varphi}/{2e},
\label{U_phi}
\end{eqnarray}
where $E_J= \hbar I_C/2e$ is the Josephson coupling energy. In order to ensure classical behavior of the
phase in the wide range of parameters we also require
\begin{eqnarray}
E_J\gg E_C,
\label{condition_EJEC}
\end{eqnarray}
where $E_C=e^2/2C$ is the charging energy.
In what follows we will focus on the regime of weak noise and require the condition
\begin{eqnarray}
k_B T^*\lesssim E_J
\label{condition_TEJ}
\end{eqnarray}
to be satisfied. This regime is the most interesting for a simple reason: 
if the environment temperature becomes comparable to or exceeds the Josephson energy,
the superconducting branch of the current-voltage characteristics becomes strongly smeared
and the heat transport properties of the Josephson junction become qualitatively similar to those of
an NIS tunnel junction, which is well investigated.

At low bias currents, $I_x<I_C$, and at sufficiently low temperatures the classical approach
based on Eq. (\ref{RSJ}) is insufficient and the quantum nature of the phase becomes important.
In this regime the phase is fluctuating around its equilibrium value
\begin{eqnarray}
\varphi_0=\arcsin(I_x/I_C),
\label{phi_0}
\end{eqnarray}
and the charge and heat transport properties of the junction may be expressed via the two
quantum phase correlation functions \cite{Ingold}
\begin{eqnarray}
\tilde {\cal P}(E) &=& e^{\left\langle (\hat\varphi(0)-\varphi_0)^2\right\rangle/2}
\int \frac{dt}{2\pi \hbar}\,e^{iEt/\hbar}\, \left\langle e^{i\hat\varphi(t)/2} e^{i\hat\varphi(0)/2}\right\rangle,
\nonumber\\  
{\cal P}(E) &=& \int \frac{dt}{2\pi \hbar}\,e^{iEt/\hbar}\, \left\langle e^{i\hat\varphi(t)/2} e^{-i\hat\varphi(0)/2}\right\rangle.
\label{PofE0}
\end{eqnarray}
These functions are expressed via the effective impedance of the
electromagnetic environment of the junction, which in our model reads
\begin{eqnarray}
Z_S(\omega)=\left(-i\omega C+\frac{1}{R_S}+\frac{2e I_C}{-i\hbar\omega}\cos\varphi_0\right)^{-1}.
\end{eqnarray}
The functions (\ref{PofE0}) may be explicitly evaluated and have the form \cite{Ingold}
\begin{eqnarray}
&& {\cal P}(E) = \theta(E_{\max}-|E|)\int \frac{dt}{2\pi \hbar} e^{iEt/\hbar+J(t)}
\label{PofE}
\\ &&
\tilde{\cal P}(E) = \theta(E_{\max}-|E|)\int \frac{dt}{2\pi \hbar} e^{iEt/\hbar-J(t)}
\label{PofE_tilde}
\\ &&
 J(t)=\frac{R_S}{R_k}\int \frac{d\omega\,\omega\left(\coth\frac{\hbar\omega}{2T^*}(\cos\omega t-1)-i\sin\omega t\right)}
 { \omega^2+R_S^2C^2\left(\omega^2-\omega_J^2\right)^2 },
\nonumber
\end{eqnarray}
where $\omega_J=\sqrt{2e I_C/\hbar C}(1-I_x^2/I_C^2)^{1/4}$ is the frequency of small oscillations around the minimum
of the Josephson potential.
Here we have also introduced an effective high energy cutoff
\begin{eqnarray}
E_{\max}\approx 2E_J\left(\sqrt{1-\frac{I_x^2}{I_C^2}}-\frac{I_x}{I_C}\left(\frac{\pi}{2}-\arcsin\frac{I_x}{I_C} \right)\right),
\label{Emax}
\end{eqnarray}
which equals to height of the barrier in the tilted Josephson potential (\ref{U_phi}).
This cutoff comes from the fact that
at higher energies the Gaussian approximation used to derive the correlation functions
(\ref{PofE},\ref{PofE_tilde}) is no longer valid. The high energy cutoff (\ref{Emax}) is important
only in the narrow range of bias currents close to $I_C$.

\section{Statistics of the heat transport}

In order to introduce the concept of the full counting statistics of the heat current
we  first define the energy ${\cal E}_1$ extracted from the superconductor 1 during a time $t$.
Since the heat transport through the junction is a stochastic
processes, the energy ${\cal E}_1$ is a random value described by a certain probability distribution ${\cal W}(t,{\cal E}_1)$. 
Our goal in this section is to find the cumulant generating function corresponding to this distribution
\begin{eqnarray}
{\cal F}(t,\chi)=\ln\left[\int d{\cal E} \,e^{-i{\cal E}_1\chi/\hbar}\,  {\cal W}(t,{\cal E}_1)\right],
\label{F00}
\end{eqnarray}
where $\chi$ is the counting field.
The function (\ref{F00}) can be expressed in the form 
\begin{eqnarray}
&&{\cal F}(t,\chi)=
\nonumber\\ &&
\ln\left[\frac{{\rm tr}\,\left[ e^{-iH_1\chi/\hbar}  e^{-iHt/\hbar} e^{iH_1\chi/\hbar} e^{-H/k_B T} e^{iHt/\hbar}\right]}{{\rm tr}\left[e^{-H/k_B T}\right]}\right].
\label{F1}
\end{eqnarray}
Provided the cumulant generating function is known,
the cooling power is given by its derivative
\begin{eqnarray}
P^{(1)} = \frac{\partial\langle {\cal E}_1\rangle}{\partial t} =  -i\hbar\frac{\partial}{\partial t}\frac{\partial {\cal F}}{\partial\chi}\bigg|_{\chi=0}. 
\label{Pdef}
\end{eqnarray}
Similarly, zero frequency spectral density of the heat current noise is defined as follows
\begin{eqnarray}
S_P^{(1)} = \frac{\partial}{\partial t}\big[\langle {\cal E}_1^2\rangle - \langle {\cal E}_1\rangle^2\big] 
=  -\hbar^2\frac{\partial}{\partial t}\frac{\partial^2 {\cal F}}{\partial\chi^2}\bigg|_{\chi=0}.
\label{SP0}
\end{eqnarray}
Both the cooling power and the noise fluctuate in time following the fluctuations of the phase $\varphi$.
Hence, in order to obtain experimentally relevant parameters one
should further average Eqs. (\ref{Pdef},\ref{SP0}) over these fluctuations.

One can derive an explicit expression for the cumulant generating function (\ref{F1}) by means
of the second order perturbation theory in the tunnel Hamiltonian $H_T$. The
details of this derivation are summarized in Appendix A. Here we directly cite
the result. The function ${\cal F}(t,\chi)$ is the sum of two contributions:
\begin{eqnarray}
{\cal F}(t,\chi)={\cal F}_{qp}(t,\chi)+{\cal F}_J(t,\chi),
\label{FGR}
\end{eqnarray}
where ${\cal F}_{qp}(t,\chi)$ and ${\cal F}_{J}(t,\chi)$ originate from, respectively, quasiparticle and Cooper pair tunneling through the junction. 
Defining the two functions characterizing the superconducting leads ($j=1,2$)
\begin{eqnarray}
W_j^{qp}(t)=\int d\epsilon_j \frac{e^{iE_j t/\hbar}f_j(E_j) +  e^{-iE_jt/\hbar}(1-f_j(E_j))}{2},
\end{eqnarray}
where $E_j=\sqrt{\epsilon_j^2+\Delta_j^2}$, and, assuming that the Josephson phase is classical, we obtain
the quasiparticle contribution in the form 
\begin{eqnarray}
{\cal F}_{qp} &=& \frac{1}{\pi\hbar  e^2 R} \int_0^t dt'\int_{-\infty}^{t'} dt''\cos\frac{\varphi(t')-\varphi(t'')}{2}
\nonumber\\ &&\times\,
\big\{ 
W_1^{qp}(t'-t''+\chi) W_2^{qp}(t'-t'')
\nonumber\\ &&
+\,W_1^{qp}(t''-t'+\chi) W_2^{qp}(t''-t')
\big\}.
\label{Fqp}
\end{eqnarray}
The Cooper pair, or Josephson contribution has a similar structure,
\begin{eqnarray}
{\cal F}_J &=& \frac{1}{\pi\hbar  e^2 R}\int_0^t dt'\int_{-\infty}^{t'} dt'' 
\cos\frac{\varphi(t')+\varphi(t'')}{2}
\nonumber\\ &&\times\,
\big\{ W^J_1(t'-t''+\chi)W^J_2(t'-t'') 
\nonumber\\ &&
+\, W^J_1(t''-t'+\chi)W^J_2(t''-t') \big\},
\label{FJ}
\end{eqnarray}
but it depends on the sum of the two phases, $\varphi(t')+\varphi(t'')$, instead of their
difference. Besides that, it contains another pair of the functions
\begin{eqnarray}
&& W^J_j(t) =  \int d\epsilon_j \frac{\Delta_j}{2 E_j}
\nonumber\\ &&\times\,
\big[ e^{iE_{j}t/\hbar} f_{j}(E_j)-e^{-iE_{j}t/\hbar} (1-f_{j}(E_j))\big].
\label{WJ}
\end{eqnarray}
The expressions (\ref{Fqp},\ref{FJ})
are valid for arbitrary time dependence of the Josephson phase $\varphi(t)$.

If the phase is quantum, the cumulant generating functions have to be
modified and take the form
\begin{eqnarray}
&& {\cal F}_{qp} = \frac{1}{ 2 \pi\hbar  e^2 R}\int_0^t dt'\int_{0}^{t'} dt''
\nonumber\\ && \times\,
\bigg\{
\left( \left\langle e^{-i\hat\varphi(t')/2}e^{i\hat\varphi(t'')/2} \right\rangle + \left\langle e^{i\hat\varphi(t')/2}e^{-i\hat\varphi(t'')/2} \right\rangle \right) 
\nonumber\\ && \times\,
W_1^{qp}(t'-t''+\chi)W_2^{qp}(t'-t'')
\nonumber\\ &&
+\,\left( \left\langle e^{-i\hat\varphi(t'')/2}e^{i\hat\varphi(t')/2} \right\rangle + \left\langle e^{i\hat\varphi(t'')/2}e^{-i\hat\varphi(t')/2} \right\rangle \right)
\nonumber\\ && \times\,
W_1^{qp}(t''-t'+\chi)W_2^{qp}(t''-t')
\bigg\},
\label{Fqp_q}
\end{eqnarray}
\begin{eqnarray}
&& {\cal F}_J =\frac{1}{ 2 \pi \hbar e^2 R}\int_0^t dt'\int_{0}^{t'} dt''
\nonumber\\ && \times\,
\bigg\{
\left( \left\langle e^{i\hat\varphi(t')/2}e^{i\hat\varphi(t'')/2} \right\rangle 
+ \left\langle e^{-i\hat\varphi(t')/2}e^{-i\hat\varphi(t'')/2} \right\rangle\right) 
\nonumber\\ && \times\,
W_1^J(t'-t''+\chi)W_2^J(t'-t'')
\nonumber\\ &&
+\,\left( \left\langle e^{i\hat\varphi(t'')/2}e^{i\hat\varphi(t')/2} \right\rangle 
+ \left\langle e^{-i\hat\varphi(t'')/2}e^{-i\hat\varphi(t')/2} \right\rangle\right) 
\nonumber\\ &&\times\,
W_1^J(t''-t'+\chi)W_2^J(t''-t')
\bigg\}.
\label{FJ_q}
\end{eqnarray}
 The angular brackets here stand for both quantum mechanical and
 statistical averaging. At bias currents below $I_C$ the fluctuations of the
 phase become Gaussian, and the average phase correlators, appearing in 
 Eqs. (\ref{Fqp_q},\ref{FJ_q}), may be expressed via the functions (\ref{PofE},\ref{PofE_tilde}).
 At $I_x>I_C$ the classical dynamics of the phase sets in, and under the conditions
 (\ref{condition_RS},\ref{condition_EJEC},\ref{condition_TEJ})
 one can replace the cumulant generating functions (\ref{Fqp_q},\ref{FJ_q}) by the classical ones (\ref{Fqp},\ref{FJ}).

\section{Cooling power}

We are now in position to evaluate the cooling power of superconductor 1. 
In the regime of classical phase fluctuations Eqs. (\ref{Pdef},\ref{Fqp},\ref{FJ})
lead to the following result
\begin{eqnarray}
P^{(1)}(t) =P_{qp}^{(1)}(t)+P_J^{(1)}(t),
\label{P1t}
\end{eqnarray}
where the quasiparticle cooling power has the form 
\begin{eqnarray}
P_{qp}^{(1)}(t) = -\frac{i}{\pi e^2 R}\int_{-\infty}^{t} dt'
\big[ \dot W_1^{qp}(t-t') W_2^{qp}(t-t')
\nonumber\\ 
+\, \dot W_1^{qp}(t'-t) W_2^{qp}(t'-t)\big]\cos\frac{\varphi(t)-\varphi(t')}{2},
\label{Pqp1}
\end{eqnarray}
and the Josephson contribution reads
\begin{eqnarray}
P_J^{(1)}(t) = -\frac{i}{\pi e^2R} \int_{-\infty}^{t} dt' 
\big[ \dot W^J_1(t-t')W^J_2(t-t') 
\nonumber\\
+\, \dot W^J_1(t'-t)W^J_2(t'-t) \big]\cos\frac{\varphi(t)+\varphi(t')}{2}. 
\label{PJ1}
\end{eqnarray}

It is interesting to compare the results (\ref{P1t},\ref{Pqp1},\ref{PJ1}) with
the well known expressions for the charge current\cite{LA,werthammer}
\begin{eqnarray}
I(t) = I_{qp}(t)+I_J(t),
\label{I1t}
\end{eqnarray}
\begin{eqnarray}
I_{qp}(t) = \frac{i}{\pi\hbar e  R}\int_{-\infty}^{t} dt'
\big\{ W_1^{qp}(t-t') W_2^{qp}(t-t') 
\nonumber\\ 
-\, W_1^{qp}(t'-t) W_2^{qp}(t'-t) \big\} \sin\frac{\varphi(t)-\varphi(t')}{2},
\label{Iqp}
\end{eqnarray}
\begin{eqnarray}
I_J(t) = \frac{i }{\pi\hbar e  R} \int_{-\infty}^{t} dt' \big\{ W^J_1(t-t')W^J_2(t-t') 
\nonumber\\
-\, W^J_1(t'-t)W^J_2(t'-t)\big\}\sin\frac{\varphi(t)+\varphi(t')}{2}.
\label{IJ}
\end{eqnarray} 
One observes two main differences between the cooling power and the charge current: 
(i) the charge current contains sines of the combinations $\varphi(t)\pm\varphi(t')$ while the cooling power -- their cosines,
and (ii) the charge current contains the functions $W_1^{qp}, W^J_1$ 
while the cooling power --- their time derivatives $\dot W_1^{qp},\dot W^J_1$.  
 
Combining Eqs. (\ref{Pqp1},\ref{PJ1}) with (\ref{Iqp},\ref{IJ}) one can easily derive
the identity reflecting the energy conservation in the junction
\begin{eqnarray}
P^{(1)}+P^{(2)}= \frac{\hbar \dot\varphi }{2e}\big(I_J-I_{qp}\big) + \frac{dA}{dt}, 
\label{conservation}
\end{eqnarray}
where $P^{(2)}$ is the cooling power of superconductor 2, which differs from $P^{(1)}$ (\ref{P1t}-\ref{PJ1})
by interchanging the indexes 1 and 2, and $A$ is a certain combination of the integrals,
which vanishes if the phase varies in time slowly. 
The last term on the right hand side of Eq. (\ref{conservation}) drops out after the averaging 
because it is a full time derivative. 
Obviously, the product $\hbar \dot\varphi I_{qp}/2e$ is the work done by the external current source. 
The term $ \hbar \dot\varphi I_{J}/2e$ in the adiabatic limit
reduces to the time derivative of the Josephson energy $-d(E_J\cos\varphi)/dt$
and vanishes upon the averaging. 
Hence the identity (\ref{conservation}) shows that the work of the current source
is, on average, split between the two superconducting leads. Moreover, if  the lead 1 is cooled and
$P^{(1)}>0$, then according to Eq. (\ref{conservation}) lead 2 is inevitably strongly heated 
because it has to absorb both the power $P^{(1)}$ and the work of the current source.
This effect is well known for NIS tunnel junctions\cite{review}.

Let us now consider a voltage biased Josephson junction and put $\varphi(t)=2eVt/\hbar$.
The cooling power takes the form
\begin{eqnarray}
P^{(1)}(t) &=& P_{qp}^{(1)}(V)+ P_{\cos}^{(1)}(V)\cos[2eVt/\hbar]
\nonumber\\ &&
+\, P_{\sin}^{(1)}(V)\sin[2eVt/\hbar].
\label{Pvt}
\end{eqnarray}
The quasiparticle contribution to it, $P_{qp}^{(1)}(V)$, becomes time independent in this case and is given by the integral\cite{FK}
\begin{eqnarray}
P_{qp}^{(1)}(V) &=& \frac{1}{ e^2 R}\int dE\,N_1(E-eV)N_2(E)
\nonumber\\ &&\times\,
(E-eV)\,\big[f_1(E-eV)-f_2(E)\big].
\label{Pqp}
\end{eqnarray}
Here we defined the quasiparticle densities of states in the leads
\begin{eqnarray}
N_j(E)={|E|\theta(|E|-\Delta_j)}\big/{\sqrt{E^2-\Delta_j^2}},\;\; j=1,2.
\end{eqnarray}
The amplitude of the "anomalous" oscillating term reads
\begin{eqnarray}
P_{\cos}^{(1)}(V) &=&  -\frac{1}{ e^2 R}\int dE\,N_1(E-eV)N_2(E)
\nonumber\\ && \times\,
\frac{\Delta_1\Delta_2}{E}\big[f_1(E-eV)-f_2(E)\big],
\label{Pcos}
\end{eqnarray}
and the Josephson amplitude has the form
\begin{eqnarray}
&& P_{\sin}^{(1)}(V) = \frac{V}{2\pi eR}   
\int d\epsilon_1 d\epsilon_2 \frac{\Delta_1\Delta_2}{E_2}\times
\nonumber\\ &&
\bigg[ \frac{1-f_1(E_1)-f_2(E_2)}{(E_1+E_2)^2-e^2V^2}  
+ \frac{f_1(E_1)-f_2(E_2)}{(E_1-E_2)^2-e^2V^2} \bigg]. 
\label{Psin}
\end{eqnarray}

\begin{figure}
\includegraphics[width=7.5cm]{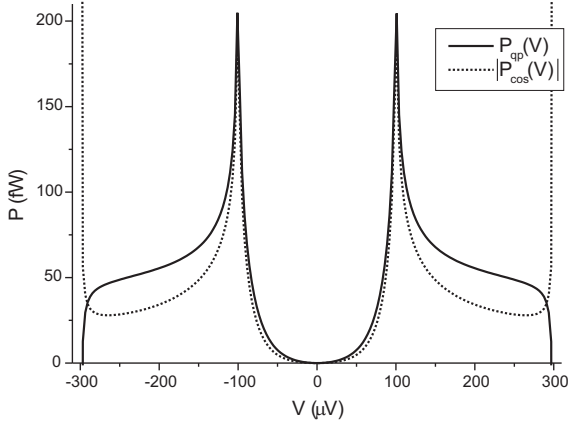}
\caption{The quasiparticle cooling power $P_{qp}(V)$ (solid line) and the absolute value of the amplitude 
$|P_{\cos}(V)|$ (dotted line). The parameters are chosen as follows: $\Delta_1=100$ $\mu$eV, $\Delta_2=200$ $\mu$eV,
$R=1$ k$\Omega$, $T_1=T_2=230$ mK. 
}
\label{fig_Pqp}
\end{figure}

\begin{figure}
\includegraphics[width=7.5cm]{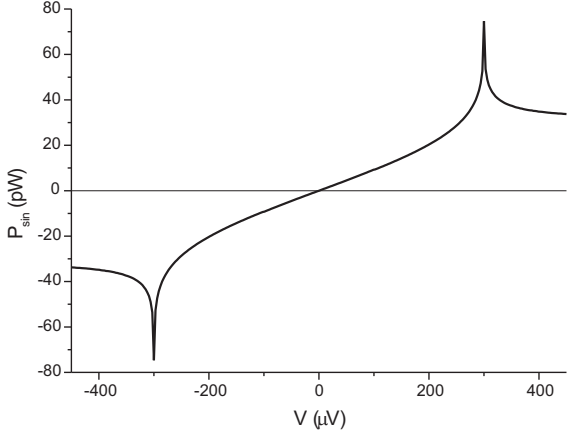}
\caption{The Josephson component of the cooling power $P_{\sin}(V)$ for the same parameters as in Fig. \ref{fig_Pqp}.}
\label{fig_Psin}
\end{figure}

The amplitudes $P_{qp}^{(1)}(V)$ and $P_{\cos}^{(1)}(V)$ are exponentially suppressed
in the limit of low bias voltage and low temperatures $eV,T_1,T_2\ll \Delta_2-\Delta_1$.
In this case we find
\begin{eqnarray}
&& P_{qp}^{(1)}(V) = \frac{\sqrt{2\pi}}{e^2R}\frac{\Delta_2^{5/2}}{\sqrt{\Delta_2^2-\Delta_1^2}}
\bigg[ \sqrt{k_B T_1}\,e^{-\Delta_2/k_B T_1}
\nonumber\\ &&\times\,
 \cosh\frac{eV}{k_B T_1} - \sqrt{k_B T_2}\, e^{-\Delta_2/k_B T_2} \bigg],
\nonumber\\
&& P_{\cos}^{(1)}(V) = -(\Delta_1/\Delta_2)P_{qp}(V).
\label{Pqp_Pcos_app}
\end{eqnarray}
In contrast to that, the Josephson amplitude $P_{\sin}^{(1)}(V)$ is not exponentially suppressed. However, it is a linear function of $V$, 
\begin{eqnarray}
P_{\sin}^{(1)}(V) = \alpha\left({\Delta_2}/{\Delta_1}\right) I_C V,
\end{eqnarray}
and tends to zero at $V=0$ for any values of the temperatures $T_1,T_2$. Here we have defined a numerical pre-factor 
\begin{eqnarray}
\alpha(\kappa) =  1-\frac{\kappa^2}{\sqrt{1-\kappa^2}}\frac{K'\left(\sqrt{1-\kappa^2}\right)}{K\left(\sqrt{1-\kappa^2}\right)} ,
\end{eqnarray}
where $K(x)$ is the complete elliptic integral of the first kind.

In the limit of zero bias, $V\to 0$, the cooling power (\ref{Pvt}) reduces to the form \cite{Guttman,Zhao1,Zhao2}
\begin{eqnarray}
P^{(1)} &=&  \frac{2}{e^2 R} \int_{\Delta_2}^\infty dE
\frac{E(E^2-\Delta_1\Delta_2\cos\varphi_0)}{\sqrt{E^2-\Delta_1^2}\sqrt{E^2-\Delta_2^2}}
\nonumber\\ &&\times\,
\big[f_1(E)-f_2(E)\big],
\label{P0}
\end{eqnarray}
and only the contributions $P^{(1)}_{qp}$ and $P^{(1)}_{\cos}$ survive.

If the voltage drop across the junction is not fixed, but the phase changes adiabatically,
$\dot\varphi\ll\min\{\Delta_1,\Delta_2,T_1,T_2\}$, one can approximate the cooling power by replacing 
$V$ by $ \hbar \dot\varphi/2e$ and $2eVt/\hbar$ by $\varphi$ in Eq. (\ref{Pvt}).
After that one arrives at the approximate expression
\begin{eqnarray}
P^{(1)}(t)&=&P^{(1)}_{qp}\left(\frac{\hbar \dot\varphi}{2e}\right)+P^{(1)}_{\cos}\left(\frac{\hbar \dot\varphi}{2e}\right)\cos\varphi
\nonumber\\ &&
+\,\alpha\left(\frac{\Delta_2}{\Delta_1}\right)\frac{\hbar \,I_C}{2e}\,\dot\varphi\sin\varphi.
\label{P1app}
\end{eqnarray}
The last Josephson term of this formula is
the full time derivative and it averages out to zero for
any realization of the phase fluctuations. 
Due to the same reason the Josephson component 
results in the contribution $\propto \omega^2$ to heat current noise,
which vanishes in zero frequency limit.  
In order to detect it one has to perform high frequency noise measurements.

The components $P^{(1)}_{qp}$ and $P^{(1)}_{\cos}$ of the cooling power are plotted in Fig. \ref{fig_Pqp} and
the Josephson component $P^{(1)}_{\sin}$ is shown in Fig. \ref{fig_Psin}.
We observe that $P^{(1)}_{qp}$ and $P^{(1)}_{\cos}$ are even functions of the bias voltage,
while the component $P^{(1)}_{\sin}$ is an odd function of it.

\section{Heat current noise}
\label{sec:Heatnoise}
Low frequency heat current noise, or the noise of the cooling power, $S_P=\int dt'\langle \delta P^{(1)}(t)\delta P^{(1)}(t')\rangle$,
can be derived from the Eqs.  (\ref{SP0},\ref{Fqp},\ref{FJ}) and has the form
\begin{eqnarray}
S_P =S_P^{qp}+S_P^J.
\label{SPdef}
\end{eqnarray}
Here
\begin{eqnarray}
&& S_P^{qp} = -\frac{R_k}{2\pi^2 R}\int_{-\infty}^{t} dt'
\big\{
\ddot W_1^{qp}(t-t') W_2^{qp}(t-t')
\nonumber\\ &&
+ \ddot W_1^{qp}(t'-t) W_2^{qp}(t'-t)
\big\}\cos\frac{\varphi(t)-\varphi(t')}{2}
\label{Sqpdef}
\end{eqnarray}
is the noise associated with quasiparticle tunneling and
\begin{eqnarray}
&& S_P^J = -\frac{R_k}{2\pi^2 R} \int_{-\infty}^{t} dt' 
\big\{ \ddot W^J_1(t-t')W^J_2(t-t') 
\nonumber\\ &&
+\, \ddot W^J_1(t'-t)W^J_2(t'-t) \big\}\cos\frac{\varphi(t)+\varphi(t')}{2}
\label{SJdef}
\end{eqnarray}
is the Josephson contribution to the noise.

Considering again a voltage biased junction with $\varphi(t)=2eVt/\hbar$, we find\cite{GP}
\begin{eqnarray}
S_P^{qp} &=& \frac{1}{e^2R}\int dE N_1(E-eV)N_2(E) (E-eV)^2 
\nonumber\\ &&\times\,
\big[ f_1(E-eV)(1-f_2(E)) 
\nonumber\\ &&
+\, (1-f_1(E-eV))f_2(E) \big].
\label{Sqp}
\end{eqnarray}
The Josephson noise in this case takes the from
\begin{eqnarray}
S_P^J = S_{\cos}(V) \cos[2eVt/\hbar] + S_{\sin}(V) \sin[2eVt/\hbar],
\end{eqnarray}
where
\begin{eqnarray}
S_{\cos}(V) &=& -\frac{1}{e^2R}\int dE\, N_1(E-eV)N_2(E) 
\nonumber\\ &&\times\,
\frac{\Delta_1\Delta_2 (E-eV)}{E}\big[ f_1(E-eV) (1-f_2(E)) 
\nonumber\\ &&
+\, (1-f_1(E-eV))f_2(E) \big],
\end{eqnarray}
and
\begin{eqnarray}
&& S_{\sin}(V) 
= -\frac{V}{2\pi eR}\int d\epsilon_1 d\epsilon_2 \frac{\Delta_1\Delta_2 E_1}{E_2} 
\nonumber\\ && \times\,
\bigg\{
\frac{f_1(E_1)f_2(E_2)+[1-f_1(E_1)][1-f_2(E_2)]}{(E_1+E_2)^2-e^2V^2}
\nonumber\\ &&
-\,\frac{f_1(E_1)[1-f_2(E_2)]+[1-f_1(E_1)]f_2(E_2)}{(E_1-E_2)^2-e^2V^2} 
\bigg\}.
\end{eqnarray}
$S_{\cos}(V)$ and $S_{\sin}(V)$ are the amplitudes
of oscillating components of the noise and unlike  $S_{qp}(V)$, which is always positive,
may change their sign at certain values of the bias voltage. For example,
the amplitude $S_{\cos}(V)$ is negative at $e|V|<\Delta_1+\Delta_2$ and positive 
otherwise.  

In a typical experimental setup the time averaged value of the noise is measured.
In this case the oscillating components vanish and the noise spectral density is given
by the quasiparticle noise $S_P^{qp}$ (\ref{Sqp}).

At zero bias voltage, $V=0$, the phase takes its equilibrium value $\varphi_0$ (\ref{phi_0}) and we get
\begin{eqnarray}
S_P &=& \frac{1}{e^2R}\int dE N_1(E)N_2(E)\, \big(E^2-\Delta_1\Delta_2\cos\varphi_0 \big) 
\nonumber\\ &&\times\,
\big[ f_1(E)(1-f_2(E)) 
+ (1-f_1(E))f_2(E) \big].
\label{SP_zero}
\end{eqnarray}
One can verify that in equilibrium, i.e. at $T_1=T_2=T$, zero bias noise satisfies the fluctuation dissipation theorem in the form
\begin{eqnarray}
S_P = 2k_B T^2 \frac{\partial P^{(1)}}{\partial T_1}\bigg|_{T_1=T_2=T},
\label{FDT}
\end{eqnarray}
where $P^{(1)}$ is given by Eq. (\ref{P0}). 

\begin{figure}
\includegraphics[width=7.5cm]{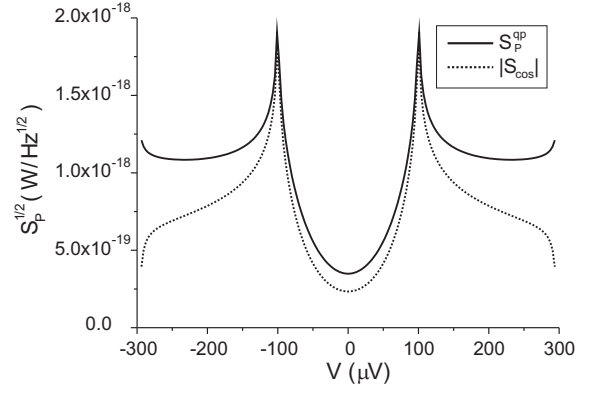}
\caption{The components of the heat current noise $S^{qp}_P(V)$ (solid line) and the  
$|S_P^{\cos}(V)|$ (dotted line).
The parameters of the junction are the same as in Fig. (\ref{fig_Pqp}).}
\label{fig_Sqp}
\end{figure}

The noise for typical values of the junction parameters is plotted in Fig. \ref{fig_Sqp}.
It exhibits characteristic peaks at bias voltages $eV=\pm(\Delta_2-\Delta_1)$
and jumps at $eV=\pm(\Delta_2+\Delta_1)$, which are also seen in the cooling power and the current.

\section{Effect of phase dynamics}

So far we have discussed a voltage biased Josephson junction and
a junction in superconducting state with a fixed phase.
In practice these regimes are not easy to achieve and, in general, one has to
consider a junction in combination with its electromagnetic environment as
depicted in Fig. \ref{fig_schematics}. The effect of the environment, and 
of the phase dynamics associated with it, on the cooling power of the junction 
will be the subject of this section. 

We begin with the regime of high bias current and assume
that the conditions (\ref{condition_RS},\ref{condition_EJEC},\ref{condition_TEJ}) are satisfied.
The phase in this case may be treated classically and obeys Eq. (\ref{RSJ}).
In order to simplify our analysis further, we will consider an overdamped Josepshon junction 
assuming that $1/R_SC\gg \sqrt{2eI_C/\hbar C}$. This condition allows one to put $C=0$ in the equation of motion (\ref{RSJ}),
which reduces it to the exactly solvable\cite{AH} resistively shunted Josephson junction model (RSJ-model).
Some details on the RSJ model needed for our analysis are briefly summarized in Appendix B.
Formally, in order to evaluate the average value of the cooling power one needs
to resolve Eq. (\ref{RSJ}) for $\varphi(t)$, substitute the result in Eqs. (\ref{Pqp1},\ref{PJ1})
and evaluate the path integral over the noise of the shunt resistor $\xi_S(t)$.
However, at high values of bias current one can avoid doing that and instead use a simple approximate approach. 
In order to justify it, we note that the integrals in Eqs. (\ref{Pqp1},\ref{PJ1}) converge if $t-t'\gtrsim \hbar/2eV(t)$,
where $V(t)= \hbar\dot\varphi(t)/2e$ is the instantaneous value of the voltage.
If the condition
\begin{eqnarray}
\frac{1}{4}\left\langle \big[\delta\varphi(t+\hbar/2eV(t))-\delta\varphi(t)\big]^2 \right\rangle_\xi\! \approx\! \frac{e R_S k_BT^*}{2 \hbar V(t)}\! \ll 1
\label{phase_dif}
\end{eqnarray} 
is satisfied, one can ignore the noise in Eq. (\ref{RSJ}) while solving it on the short time intervals of the duration $\sim \hbar/2eV(t)$. 
Then from Eq. (\ref{RSJ}) one finds 
\begin{eqnarray}
V(t)=R_S(I_x-I_C\sin\varphi(t)),
\label{replacement}
\end{eqnarray} 
and the voltage is restricted to the interval 
\begin{eqnarray}
R_S(I_x-I_C)<V(t)<R_S(I_x+I_C).
\label{interval}
\end{eqnarray} 
The condition (\ref{phase_dif}) then translates into a simpler one 
\begin{eqnarray}
I_x-I_C\gg \frac{e}{\hbar}k_BT^*.
\label{condition1}
\end{eqnarray}
Thus, provided this condition is satisfied, one can make the replacement (\ref{replacement})
and perform the noise averaging in Eqs. (\ref{P1t}-\ref{IJ})
with the aid of the following approximate formula  
\begin{eqnarray}
&& \left\langle e^{i[\varphi(t)-\varphi(t')]/2}  \right\rangle_{\xi_S} \approx 
\nonumber\\ &&
\int d\varphi\,e^{ieR_S(I_x-I_C\sin\varphi)(t-t')/\hbar}\sigma(\varphi).
\end{eqnarray}  
Here $\sigma(\varphi)$\cite{AH} is the distribution function of the phase (\ref{W0}). 
As a result, the average current (\ref{I1t},\ref{Iqp},\ref{IJ}) and the cooling power (\ref{P1t},\ref{Pqp1},\ref{PJ1}) acquire the form
\begin{eqnarray}
\langle I\rangle &=& \int_{-\pi}^\pi d\varphi \sigma(\varphi)
\big[ I_{qp}( I_xR_S - I_CR_S\sin\varphi) 
\nonumber\\ &&
+\, I_C\sin\varphi \big],
\label{Iav}
\\
\langle P^{(1)}\rangle &=& \int_{-\pi}^\pi d\varphi \sigma(\varphi)
\big[ P_{qp}( I_xR_S - I_CR_S\sin\varphi) 
\nonumber\\ &&
+\, P_{\cos}( I_xR_S - I_CR_S\sin\varphi) \cos\varphi \big],
\label{Pav}
\end{eqnarray}
where 

\begin{eqnarray}
I_{qp}(V) = \int dE \frac{N_1(E-eV)N_2(E)}{eR}
\big[f_1(E-eV)-f_2(E)\big]
\nonumber
\end{eqnarray}
is the usual quasiparticle current.

\begin{figure}
\includegraphics[width=7.5cm]{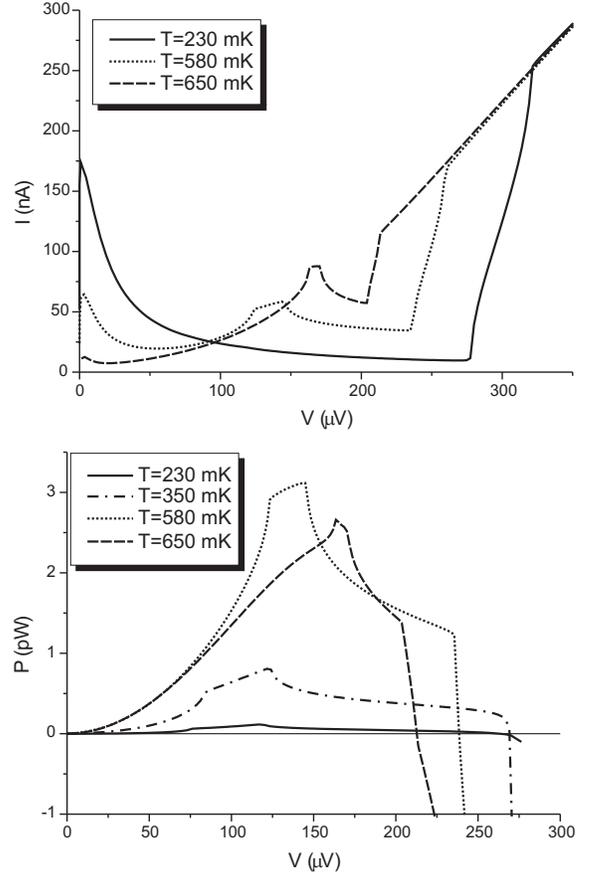}
\caption{The time averaged current $\langle I\rangle$ (\ref{Iav}) (top panel) and cooling power $\langle P\rangle$ (\ref{Pav}) (lower panel)
at different temperatures. 
The parameters are the same as in Fig. \ref{fig_Pqp}, i.e. $\Delta_1(0)=100$ $\mu$eV, $\Delta_2(0)=200$ $\mu$eV,
$R=1$ k$\Omega$. The temperature dependence of the energy gaps $\Delta_1(T),\Delta_2(T)$ is  
modeled by Bardeen-Cooper-Schrieffer  theory.
Besides that we have put $T_1=T_2=T^*$ and chosen $R_S=100$ $\Omega$.
The horizontal axis in both plots shows the average voltage drop across the junction $\langle V\rangle$
given by Eq. (\ref{Vav}).}
\label{fig_PI}
\end{figure}

The current (\ref{Iav}) and the cooling power (\ref{Pav}) are plotted in Fig. \ref{fig_PI}
versus the average voltage drop across the junction $\langle V\rangle$ (\ref{Vav}).  
As expected, strong noise of the shunt resistor suppresses the superconducting branch of the I-V curve at high temperatures.
At the same time, it does not drastically change the cooling power as compared to
the voltage biased case shown in Fig. \ref{fig_Pqp}.
The only effect of the noise at high bias (\ref{condition1}) is the smearing
of the peaks located at $eV=\Delta_2-\Delta_1$. This smearing is caused by the
fluctuations of the voltage in the interval (\ref{interval}).

Next, we consider a simple example of an  overdamped junction shunted by a noiseless resistor and put $\xi_S=0$ and $C=0$ in the
dynamical equation (\ref{RSJ}). This regime is formally achieved in the limit  $1/R_SC\gg \sqrt{2e I_C/\hbar C}$, $k_B T^*/E_J\to 0$ and $e^2R_S/\hbar\to 0$.
The model becomes exactly solvable in this case because the condition (\ref{condition1})
is always satisfied at $I_x>I_C$, and at $I_x<I_C$ the phase is just pinned to the equilibrium value $\varphi_0$ (\ref{phi_0}). 
Hence at $I_x<I_C$  the cooling power  is  given by Eq. (\ref{P0}).
At $I_x>I_C$ the Eq. (\ref{RSJ}) can be solved exactly and one finds (\cite{AL_JETP})
\begin{eqnarray}
V(t)=\frac{\hbar \dot\varphi(t)}{2e} = \frac{(I_x^2-I_C^2)R_S}{I_x-I_C\sin\omega_0 t}.
\label{V(t)}
\end{eqnarray}
Here $\omega_0=2eR_S\sqrt{I_x^2-I_C^2}/\hbar$ is the Josephson frequency 
proportional to the average voltage $\langle V\rangle =R_S\sqrt{I_x^2-I_C^2}$.
The average cooling power is found by the time averaging of the Eq. (\ref{P1app}) and reads
\begin{eqnarray}
\langle P^{(1)}\rangle = \frac{\omega_0}{2\pi}\int_{-\pi/\omega_0}^{+\pi/\omega_0} dt 
\big\{ P_{qp}^{(1)}(V) + \cos\varphi P_{\cos}^{(1)}(V) \big\}.
\label{Pav00}
\end{eqnarray}
One can verify that in the limit $k_B T^*/E_J\to 0$ the Eq. (\ref{Pav00}) matches Eq. (\ref{Pav})
because in this limit the phase distribution function (\ref{W0}) reduces to the form
$\sigma(\varphi)=\sqrt{I_x^2-I_C^2}/2\pi (I_x-I_C\sin\varphi)$ for $I_x>I_C$
and $\sigma(\varphi)=\delta\left(\varphi-\arcsin(I_x/I_C)\right)$ for $I_x<I_C$.

\begin{figure}
\begin{center}
\includegraphics[width=7.5cm]{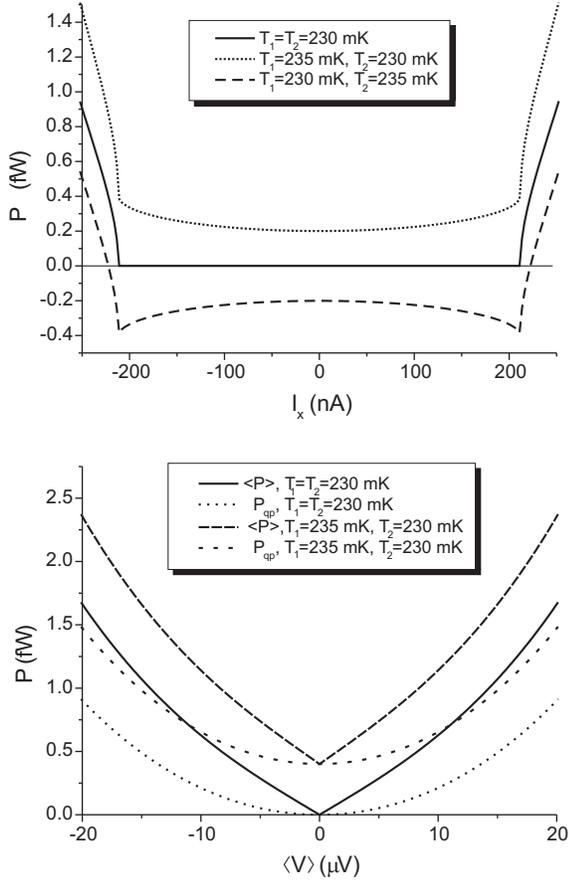}
\end{center}
\caption{The average cooling power $\langle P^{(1)}\rangle$ (\ref{P0},\ref{Pav00}) of a resistively shunted Josephson junction at low bias
and in the absence of the noise.
Top panel: $\langle P^{(1)}\rangle$ as a function of the bias current  at $T_1=T_2$, $T_1>T_2$ and $T_1<T_2$.
Lower panel: $\langle P^{(1)}\rangle$ as a function of the time averaged voltage $\langle V\rangle = R_S\sqrt{I_x^2-I_C^2}$;
for comparison we have also shown the quasiparticle cooling power $P_{qp}$ (\ref{Pqp_Pcos_app}) of a voltage biased Josephson junction.
The parameters are chosen as follows: $\Delta_1=100$ $\mu$eV, $\Delta_2=200$ $\mu$eV, $R=1$ k$\Omega$, $R_S=0.1$ k$\Omega$.
For these parameters we find $I_C=212$ nA and $I_CR_S=21.2$ $\mu$V.}
\label{fig_Plb}
\end{figure}

The time dependence of the $\cos\varphi$ at $I_x>I_C$ reads
\begin{eqnarray}
\cos\varphi = \sqrt{I_x^2-I_C^2}\cos\omega_0t\big/(I_x-I_C\sin\omega_0t).
\end{eqnarray}
With the aid of this result one can show that
the term $\propto P_{\cos}$ in Eq. (\ref{Pav00}) 
identically vanishes. 

In the most interesting low bias and low temperature regime, $eV\lesssim k_B T_1,T_2\ll\Delta_1,\Delta_2$, 
we can approximate the quasiparticle cooling power by a simple formula which follows
from the Eq. (\ref{Pqp_Pcos_app}),
\begin{eqnarray}
&&  P^{(1)}_{qp}(V)      
\approx \, \frac{\sqrt{2\pi}}{e^2R}\frac{\Delta_2^{5/2}}{\sqrt{\Delta_2^2-\Delta_1^2}} \times\\
&& \nonumber \left[\! \sqrt{k_BT_1}\, e^{-\frac{\Delta_2}{k_BT_1}}\!\left(\!1\!+\!\frac{1}{2}\! \left( \frac{eV}{k_B T_1}\right) ^2 \right)\! -\! \sqrt{k_B T_2}\, e^{-\frac{\Delta_2}{k_B T_2}} \right]
\label{Ptotal}
\end{eqnarray}
Afterwards, the time integral in Eq. (\ref{Pav00}) is easily evaluated, and at $I_x>I_C$ one arrives at the result
\begin{eqnarray}
&& \langle P^{(1)}\rangle = \delta P(\langle V\rangle) 
\nonumber\\ &&
+\,
\frac{\sqrt{2\pi}}{e^2R}\frac{\Delta_2^{5/2}\left[ \sqrt{k_B T_1}\, e^{-\frac{\Delta_2}{ k_BT_1}} - \sqrt{k_BT_2}\, e^{-\frac{\Delta_2}{k_BT_2}} \right]}{\sqrt{\Delta_2^2-\Delta_1^2}}.
\end{eqnarray}
Here the contribution 
\begin{eqnarray}
\delta P(\langle V\rangle) = \sqrt{\frac{\pi}{2}} \frac{\Delta_2^{5/2}e^{-\Delta_2/k_BT_1}|\langle V\rangle|}{(k_B T_1)^{3/2}\sqrt{\Delta_2^2-\Delta_1^2}}
\frac{\sqrt{I_C^2R_S^2+\langle V\rangle^2}}{R}
\label{Pps}
\end{eqnarray}
arises from voltage oscillations (\ref{V(t)}).

It is clear from Eq. (\ref{Pps}) that at low voltage, $\langle V\rangle\lesssim I_CR_S$ the cooling power grows linearly with the voltage, 
$\delta P\propto |\langle V\rangle|$.
This unusual behavior may be understood if one recalls that phase slips are responsible for the transport in this regime. 
Indeed, at $\langle V\rangle\lesssim I_CR_S$ the time dependence of the voltage (\ref{V(t)})
reduces to a series of well separated pulses corresponding to a sequence of phase slips.   
Hence one can express the average cooling power as a product of the phase slip rate, which is robustly 
related to the average voltage as $\Gamma_{ps}=e|\langle V\rangle|/\hbar\pi$ by Josephson relation, and the energy carried away from superconductor 1 by
a single phase slip $E_{ps}$, 
\begin{eqnarray}
\delta P(\langle V\rangle) = E_{ps}\Gamma_{ps}.
\label{PEps}
\end{eqnarray}  
Comparing equations (\ref{Pps}) and (\ref{PEps}) we find
\begin{eqnarray}
&& E_{ps} \equiv \int_{-\pi/\omega_0}^{\pi/\omega_0} P_{qp}^{(1)}(V(t))dt
\label{Eps0}
\\ &&
= \frac{\pi^{3/2}}{\sqrt{2}} \frac{\hbar\Delta_2^{5/2}e^{-\Delta_2/k_BT_1}}{(k_B T_1)^{3/2}\sqrt{\Delta_2^2-\Delta_1^2}}
\frac{\sqrt{I_C^2R_S^2+\langle V\rangle^2}}{eR}
\label{Eps}\\ &&
=\frac{\pi^{3/2}}{\sqrt{2}} \frac{R_S}{R}\frac{\Delta_2^{5/2}e^{-\Delta_2/k_B T_1}}{(k_B T_1)^{3/2}\sqrt{\Delta_2^2-\Delta_1^2}}\frac{\hbar|I_x|}{e}.
\label{Eps1}
\end{eqnarray}
The energy $E_{ps}$ becomes voltage independent for $\langle V\rangle\lesssim I_CR_S$
because the phase slips become well separated. Hence  Eq. (\ref{PEps}) always
leads to the linear dependence of the cooling power on voltage in this limit.
This conclusion holds for more general models of Josephson dynamics.
For example, weak noise, $k_B T^*\ll E_J$, or finite junction capacitance do not change it.

The dependence of the cooling power of a noiseless Josephson junction on the bias current $I_x$ and
average voltage $\langle V\rangle$ is shown in  Fig. \ref{fig_Plb}.

In the remaining part of this section we will generalize the model introduced above and include 
the weak noise with $k_B T^*\lesssim E_J$ into the analysis.
Two types of phase trajectories may be distinguished in this case: phase slips and
Gaussian fluctuations around a minimum of the Josephson potential $U(\varphi)$ (\ref{U_phi}). 
Accordingly, the cooling power
may be expressed as a sum of two terms
\begin{eqnarray}
\langle P^{(1)}\rangle = P_{ps} +  P_{gs},
\label{Pnoise_lb}
\end{eqnarray}
where $P_{ps}$ is the contribution of the phase slips and $P_{gs}$ is the contribution of the Gaussian fluctuations.

Let us first consider the phase slip contribution. As we have discussed above, it is given by the product 
\begin{eqnarray}
P_{ps}=\frac{e|\langle V\rangle|}{\pi \hbar }E_{ps}.
\label{Pps2}
\end{eqnarray}
In the presence of noise the average voltage is given by Eq. (\ref{Vav}) while
the phase slip energy at $|I_x|>I_C$ is still defined by  Eqs. (\ref{Eps}) or (\ref{Eps1}).  
At $|I_x|<I_C$ every phase slip is accompanied by a voltage pulse of the form
\begin{eqnarray}
V_{ps}(t)=\frac{R_S(I_C^2-I_x^2)}{I_C\cosh\left(2eR_S\sqrt{I_C^2-I_x^2}\,t /\hbar \right)-I_x},
\end{eqnarray}
which may be derived for Eq. (\ref{RSJ}) with $C=0$ and $\xi_S=0$.
The corresponding phase slip trajectory, $\varphi_{ps}(t)=\int_0^t dt' 2eV_{ps}(t')/\hbar$, connects the
maximum, $-\pi-\varphi_0$, and the minimum, $\varphi_0$, of the Josephson potential. 
The phase slip cooling energy then takes the form
\begin{eqnarray}
&& E_{ps} = \int_{-\infty}^{\infty}\! dtP_{qp}^{(1)}(V_{ps}(t))
\nonumber\\ &&
= \sqrt{\frac{\pi}{2}}\frac{\hbar R_S \Delta_2^{5/2}\,e^{-{\Delta_2}/{k_BT_1}}}{R (k_B T_1)^{3/2}\sqrt{\Delta_2^2-\Delta_1^2}}
\nonumber\\ &&\times\,
\frac{}{}
\left(\frac{\sqrt{I_C^2-I_x^2}}{e}+\frac{2I_x}{e}\arctan\sqrt{\frac{I_C+I_x}{I_C-I_x}}\right),
\label{Eps2}
\end{eqnarray} 
which exactly matches the expression (\ref{Eps1}) at $I_x=I_C$.

Let us now derive the contribution of Gausian fluctuations $P_{gs}$.
Provided the subgap quasiparticle resistance of the junction exceeds
the quantum resistance, $R e^{\Delta_2/k_B T_1}\gg R_k$, which is often the case in experiments with small area junctions, 
we may evaluate $P_{gs}$ applying the theory of dynamical Coulomb blockade\cite{Ingold}. 
In this way we arrive at the result 
\begin{eqnarray}
P_{gs} &=& 
\frac{1}{e^2R}\int dE_1dE_2 N_1(E_1)N_2(E_2) E_1
\nonumber\\ &&\times\,
\big\{ f_1(E_1)[1-f_2(E_2)]{\cal P}(E_1-E_2) 
\nonumber\\ &&
-\, [1-f_1(E_1)]f_2(E_2){\cal P}(E_2-E_1) \big\}
\nonumber\\&&
-\,\frac{e^{-\langle\delta\varphi^2\rangle/2}\cos\varphi_0}{e^2R}
\int dE_1dE_2 N_1(E_1)N_2(E_2)
\nonumber\\ &&\times\,
\frac{\Delta_1\Delta_2}{E_2}
\big\{ f_1(E_1)[1-f_2(E_2)]\tilde {\cal P}(E_1-E_2) 
\nonumber\\ &&
-\, [1-f_1(E_1)]f_2(E_2)\tilde {\cal P}(E_2-E_1) \big\}.
\label{Pav_min}
\end{eqnarray} 
Here the we have introduced the average square of the phase fluctuations 
\begin{eqnarray}
\frac{\langle\delta\varphi^2\rangle}{2}=\frac{e^2R_S}{\pi \hbar }\int_0^\infty d\omega
\frac{\omega\coth\frac{\hbar\omega}{2k_B T^*}}{\omega^2+R_S^2C^2\left(\omega^2-\omega_J^2\right)^2},
\end{eqnarray}
and the functions ${\cal P}(E),\tilde{\cal P}(E)$  
are defined in Eqs. (\ref{PofE},\ref{PofE_tilde}).
They obey the detailed balance principle
\begin{eqnarray}
{\cal P}(-E)=e^{E/k_B T^*}{\cal P}(E),\;\; \tilde{\cal P}(-E)=e^{E/k_B T^*}\tilde{\cal P}(E),
\end{eqnarray}
which ensures that $P_{gs}\equiv 0$ at $T_1=T_2=T^*$, as required by thermodynamics. We also note that in the limit $R_S\to 0$
one finds ${\cal P}(E)=\tilde{\cal P}(E)=\delta(E)$, $\langle\delta\varphi^2\rangle=0$ and
the expression (\ref{Pav_min}) reduces to Eq.  (\ref{P0}) as expected.

In the limit $R_S\ll R_k$ one can expand the function ${\cal P}(E)$ (\ref{PofE}) in powers of $J(t)$, which leads
to the approximate expression
\begin{eqnarray}
{\cal P}(E) &=& 
\frac{e^2R_S}{\pi\hbar}\frac{\theta(E_{\max}-|E|)}{E^2+R_S^2C^2\left( E^2-\hbar^2\omega_J^2 \right)^2}
\nonumber\\ &&\times\,
\frac{E}{1-e^{-E/k_B T^*}}
+\left(1-\frac{\langle\delta\varphi^2\rangle}{2}\right)\delta(E).
\end{eqnarray}
The function $\tilde{\cal P}(E)$ in this limit differs  from ${\cal P}(E)$ only by the sign in front of the first term. 
Within this approximation and for equal temperatures of the superconducting leads, $T_1=T_2=T$, 
the Gaussian contribution to the cooling power (\ref{Pav_min})
takes the form
\begin{eqnarray}
P_{gs} &=& \frac{2e^2R_S}{\pi \hbar}\int_0^{E_{\max}} dE 
\bigg[P_{qp}\left(\frac{E}{e}\right) - e^{-\frac{\langle\delta\varphi^2\rangle}{2}}\cos\varphi_0
\nonumber\\ &&\times\,
 P_{\cos}\left(\frac{E}{e}\right)\bigg] \frac{1}{E^2+R_S^2C^2\left( E^2-\hbar^2 \omega_J^2 \right)^2}
\nonumber\\ &&\times\,
\left(\frac{E}{e^{E/k_B T^*}-1}-\frac{E}{e^{E/k_BT}-1}\right).
\label{Pgs_app}
\end{eqnarray}

\begin{figure}
\begin{center}
\includegraphics[width=7.5cm]{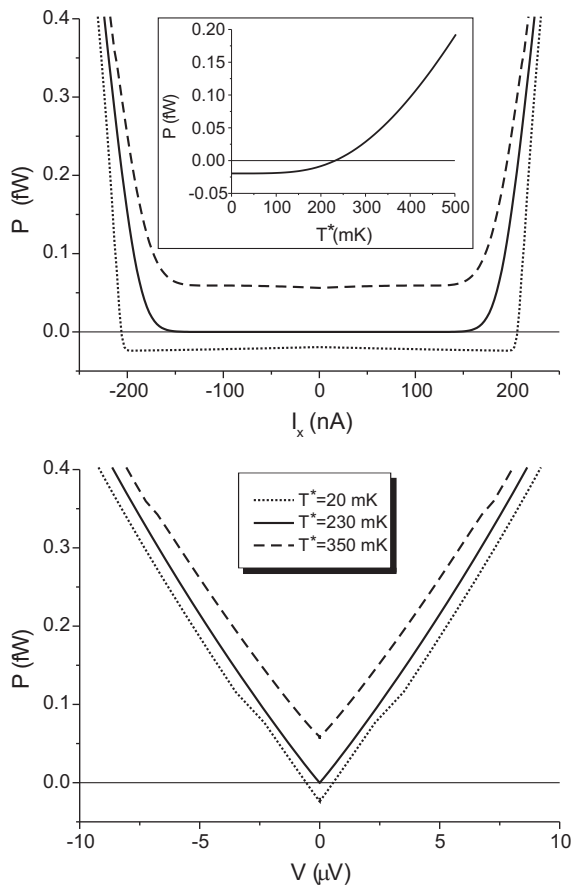}
\end{center}
\caption{Time averaged cooling power $\langle P\rangle$ (\ref{Pnoise_lb},\ref{Pps2},\ref{Pgs_app}) at $T_1=T_2=230$ mK and at
various values of the noise temperature $T^*$. Other parameters are the same as in Fig. \ref{fig_Plb}.
Top panel: Cooling power as a function of the bias current, inset: zero bias cooling power versus the environment temperature $T^*$.
Bottom panel: Cooling power as a function of the function of the average voltage for three different values of $T^*$. }
\label{fig_P_noise}
\end{figure}

The average cooling power affected by the noise (\ref{Pnoise_lb},\ref{Pps2},\ref{Pgs_app})
is plotted in Fig. \ref{fig_P_noise}. Comparing it with the cooling power of the noiseless junction, Fig. \ref{fig_Plb},
we observe two major differences. The first one is an expected smearing of the sharp features 
of the $\langle P\rangle (I_x)$ dependence by the noise. The second difference becomes visible
at $|I_x|<I_C$ and  $T_1=T_2=T$, where the cooling power vanishes in the absence of the noise but
remains finite if the noise is present.
The noise induced cooling power is positive if $T^*>T$ and negative in the opposite case $T^*<T$.
Thus, even at zero bias current one can cool the superconductor 1 by heating the shunt resistor  
and heat it by cooling the resistor down, see the inset in the top panel of Fig. \ref{fig_P_noise}.
This interesting effect has been first predicted for an NIS tunnel junction\cite{PH}. 

Finally we briefly note that although phase dynamics significantly modifies the 
average cooling power at low bias, it only weakly affects the heat current noise.
Essentially, one can always use the equations (\ref{Sqp},\ref{SP_zero}) to make estimates of the noise.
The reasons for that are simple. First, the energy carried by one quasiparticle
and the rate of quasiparticle tunneling in most cases exceed, respectively, the energy carried
by one phase slip and the rate of phase slip tunneling. Second, the noise
remains finite even at zero bias and at $T_1=T_2=T^*$. Thus its averaging over phase
motion may only lead to a certain smearing of the  bias dependence of the noise without
qualitatively changing it.

With the help of sections \ref{sec:Heatnoise} and VI, we can predict the characteristics of an electronic
 thermometer based on a Josephson junction between two different superconductors. 
 To make a simple estimate of the sensitivity and the noise properties, let us consider here a small superconducting island  of volume $V=100 \text{ nm} \times 50 \text{ nm}\times 20 \text{ nm} $ and with energy gap $\Delta_2$. 
 This island is connected to a superconducting lead with gap $\Delta_1$  by a tunnel junction with normal resistance $R_n$.
The temperature dependence of the zero bias resistance can be used as a thermometer,
\begin{eqnarray}
R_0(T_2)=\left ( \frac{\partial \langle V \rangle}{\partial \langle I \rangle }\right ) _{I_x=0} = \frac{R_{qp}(T_2)}{I_0^2(\frac{\hbar I_c(T_2)}{2 e k_B T_2})}.
\label{R0}
\end{eqnarray}

Eq. (\ref{R0}) is derived from (\ref{Iav}) and (\ref{Vav}) assuming $R_{qp}>>R_s$. One can achieve a sensitivity of $2.3 \text{ k} \Omega/\text{K}$ using the parameters given in Fig. \ref{fig_Pqp}.
In order to evaluate the performance of such a thermometer, two aspects might be considered. Both are derived from the heat balance equation applied to the island :
\begin{eqnarray}
C_e \frac{dT_2}{dt}=P^{(1)}(t)+P_{e-ph}(T_2)
\label{heateq}
\end{eqnarray}
We include here the electron-phonon heat transfer\cite{AT}  $P_{e-ph}$ 
 to balance the heat $P^{(1)}(t)$ generated by the junction. We can then verify that the thermometer does not perturb the island temperature even if the heat capacity 
$\frac{C_e}{\gamma Tc}\approx 8.5 \exp^{(-1.44 T_c/T)}$ 
of the superconductor vanishes exponentially at low temperatures, according to the BCS theory. Because $\langle P^{(1)}\rangle$ is also suppressed exponentially in the small voltage limit, 
the average junction heating is negligible.
Nevertheless, using the noise properties derived in section V allows
one to expect a noise temperature spectrum given by (\ref{FDT}) and the Fourier transform of (\ref{heateq}) in the limit of small fluctuations
\begin{eqnarray}
S_T(\omega)=\frac{2k_BT^2}{\frac{\partial P^{(1)}}{\partial T}+\frac{\partial P_{e-ph}}{\partial T}} \frac{1}{1+(\omega/\omega_c)^2}.
\end{eqnarray}
 In the low frequency limit $\omega << \omega_c =( \frac{\partial P^{(1)}}{\partial T}+\frac{\partial P_{e-ph}}{\partial T})/C_e $, 
 one can estimate the zero frequency noise of the island temperature to be $S^{1/2}_T(0)=4.0 \times 10^{-6} \text{ K}/\sqrt{\text{Hz}}$ at $230 \text{ mK}$.

\section{Summary}
In summary, we have analyzed the heat transport through a Josephson tunnel junction.
We have developed the full counting statistics approach to this problem and derived
general expressions for the cumulant generating function of the heat extracted from one of the
superconducting leads (\ref{FGR}-\ref{FJ}),
for the corresponding cooling power (\ref{P1t}-\ref{PJ1}) and for the heat
current noise (\ref{SPdef}-\ref{SJdef}). These general expressions are valid for an arbitrary 
time dependence of the Josephson phase. In analogy with the charge current, the cooling power
is the sum of the quasiparticle and Josephson contributions. The latter contribution oscillates
in time and averages to zero if the junction is biased above the critical current. 
We have also generalized all these results
to the regime where the quantum nature of the Josephson phase becomes important.
Combining these findings with the conventional theory of the Josephson phase dynamics, we have
derived the average cooling power of the junction, which can be measured in the
experiment. We find, in particular, that at low bias the average cooling power is proportional
to the voltage drop on the junction. Finally, we have also shown that zero bias cooling power
of the junction may be tuned by the heating or cooling the resistor connected in parallel to it.

The authors wish to thank Academy of Finland Centre of Excellence and Q-NET project for financial support.

\appendix

\section{Explicit expression for the generating function}

In this appendix we derive an explicit expression for  
the cumulant generating function ${\cal F}(t,\chi)$ (\ref{F1}) by means of the perturbation theory 
in tunnel Hamiltonian. The first step of our derivation is to transform ${\cal F}(t,\chi)$ to the form
\begin{eqnarray}
{\cal F}(t,\chi)=\ln\left[\frac{{\rm tr}\,\left[  e^{-iH_{\chi} t/\hbar}  e^{-H/k_B T} e^{iHt/\hbar}\right]}{\big/{\rm tr}\left[e^{-H/k_B T}\right]}\right],
\label{Z1}
\end{eqnarray}
where we have introduced the transformed Hamiltonian
\begin{eqnarray}
H_{\chi} = e^{-iH_1\chi/\hbar}  H  e^{iH_1\chi/\hbar}. 
\end{eqnarray}
Since the Hamiltonians $H_1$ and $H_2$ commute with  $H_1$, only the tunnel Hamiltonian changes
under this transformation. Thus we get
\begin{eqnarray}
H_{\chi} = H_{1} + H_{2}  + H_T(\chi),
\end{eqnarray} 
where
\begin{eqnarray}
&& H_T(\chi) = e^{-iH_1\chi/\hbar} H_T e^{iH_1\chi/\hbar}
\nonumber\\ &&
=\, \sum_{\sigma}\sum_{kn} \big[ t_{kn} e^{i\hat\varphi/2} a^\dagger_{\sigma n} c_{\sigma k}(\chi)
+t_{kn}^* e^{-i\hat\varphi/2}c^\dagger_{\sigma k}(\chi) a_{\sigma n} \big].
\nonumber\\
\label{HTxi}
\end{eqnarray}
The operators $c_{\sigma k}(\chi),c_{\sigma k}^\dagger(\chi)$ in this expression are defined as follows
\begin{eqnarray}
c_{\sigma k}(\chi) = e^{-iH_1\chi/\hbar}\, c_{\sigma k}\, e^{iH_1\chi/\hbar},
\nonumber\\
c_{\sigma k}^\dagger(\chi) = e^{-iH_1\chi/\hbar}\, c_{\sigma k}^\dagger\, e^{iH_1\chi/\hbar}.
\nonumber
\end{eqnarray}

Next we perform  Bogolubov transformation of the operators:
\begin{eqnarray}
c_{\uparrow k} = u_{1k} \gamma_{\uparrow k} - v_{1k} \gamma^\dagger_{\downarrow -k},\;\;\;
c_{\downarrow k} = v_{1k} \gamma_{\uparrow -k}^\dagger + u_{1k} \gamma_{\downarrow k},
\nonumber\\
a_{\uparrow k} = u_{2k} \alpha_{\uparrow k} - v_{2k} \alpha^\dagger_{\downarrow -k},\;\;\;
a_{\downarrow k} = v_{2k} \alpha_{\uparrow -k}^\dagger + u_{2k} \alpha_{\downarrow k}.
\label{Bogolubov}
\end{eqnarray}
Here $\gamma,\gamma^\dagger$ are the quasiparticle annihilation and creation operators in
lead 1 and $\alpha,\alpha^\dagger$ are similar operators acting in lead 2. Besides that
we have introduced the coherence factors (here $E_{jk}=\sqrt{\epsilon_{jk}^2+\Delta_j^2}$)
\begin{eqnarray}
u_{jk}=\frac{1}{\sqrt{2}}\sqrt{1+\frac{\epsilon_{jk}}{E_{jk}}},\; v_{jk}=\frac{1}{\sqrt{2}}\sqrt{1-\frac{\epsilon_{jk}}{E_{jk}}}. 
\label{uv}
\end{eqnarray}
The averaging of the products of the quasiparticle operators results in the 
distribution functions in the leads
\begin{eqnarray}
\langle\gamma^\dagger_{\sigma k}\gamma_{\sigma k}\rangle=f_{1}(E_{1k}),\;\;
\langle \alpha^\dagger_{\sigma n}\alpha_{\sigma n} \rangle=f_{2}(E_{2k}).
\label{gamma_av}
\end{eqnarray}
After the transformation (\ref{Bogolubov}) the Hamiltonian $H_1$ acquires the diagonal form
\begin{eqnarray}
H_1 = \sum_k(\epsilon_{1k}-E_{1k})+\sum_k E_{1k}\big[\gamma_{\uparrow k}^\dagger \gamma_{\uparrow k} + \gamma_{\downarrow k}^\dagger \gamma_{\downarrow k}\big].
\end{eqnarray}

The next step is two switch to the Heisenberg representation. In order to do that we split the total Hamiltonian
into two parts: the exactly solvable part $H_0=H_1+H_2$, which includes only the Hamiltonians of the leads,
and the perturbation $V_{\chi}\equiv H_T(\chi)$. Next we introduce the time ordered exponent $S_\chi(t)$,
\begin{eqnarray}
&& e^{-iH_{\chi} t/\hbar} = e^{-iH_0 t/\hbar} S_{\chi}(t), 
\nonumber\\ &&
S_{\chi}(t) = T\exp\left[ -\frac{i}{\hbar}\int_0^t dt' V_{\chi}(t') \right],
\nonumber
\end{eqnarray}
where we defined the time dependent tunnel Hamiltonian
\begin{eqnarray}
V_{\chi}(t) &=& e^{iH_0t/\hbar}H_T(\chi)e^{-iH_0t/\hbar}  
\nonumber\\
&=& \sum_\sigma\sum_{kn} \big[ t_{kn} e^{i\hat\varphi(t)/2} a^\dagger_{\sigma n}(t) c_{\sigma k}(t,\chi) 
\nonumber\\ &&
+\, t_{kn}^* e^{-i\hat\varphi(t)/2}c^\dagger_{\sigma k}(t,\chi) a_{\sigma n}(t) \big].
\label{Vt}
\end{eqnarray}
The time dependent creation and annihilation operators in this expression read
\begin{eqnarray}
a_{\uparrow n}(t) &=& u_{2n} e^{-iE_{2n}t/\hbar} \alpha_{\uparrow k} - v_{2n} e^{iE_{2n}t/\hbar} \alpha^\dagger_{\downarrow -n},
\nonumber\\
a_{\downarrow n}(t) &=& u_{2n} e^{-iE_{2n}t/\hbar} \alpha_{\downarrow k} + v_{2n} e^{iE_{2n}t/\hbar} \alpha^\dagger_{\uparrow -n};
\nonumber\\
c_{\uparrow k}(t,\chi) &=& u_{1k} e^{iE_{1k}\chi/\hbar} e^{-iE_{1k}t/\hbar} \gamma_{\uparrow k} 
\nonumber\\ &&
-\, v_{1k} e^{-iE_{1k}\chi/\hbar} e^{iE_{1k}t/\hbar} \gamma^\dagger_{\downarrow -k},
\nonumber\\
c_{\downarrow k}(t,\chi) &=& u_{1k} e^{iE_{1k}\chi/\hbar} e^{-iE_{1k}t/\hbar} \gamma_{\downarrow k} 
\nonumber\\ &&
+\, v_{1k} e^{-iE_{1k}\chi/\hbar} e^{iE_{1k}t/\hbar} \gamma^\dagger_{\uparrow -k}.
\label{c_xit}
\end{eqnarray}

After these transformations the cumulant generating function (\ref{Z1}) takes the form
\begin{eqnarray}
{\cal F}(t,\chi)=\ln\left[{\rm tr}\,\left[  S_{\chi}(t) e^{-H/k_B T} S_0^\dagger (t)\right]\big/{\rm tr}\left[e^{-H/k_B T}\right]\right].
\nonumber
\end{eqnarray}
Performing a formal expansion of the operator $S_\chi(t)$ in this expression up to the second order in $V_\chi$ we find
\begin{eqnarray}
{\cal F}&=&\int_0^t dt'\int_{0}^{t'} dt'' \big[ -\langle V_{\chi}(t')V_{\chi}(t'') \rangle  
- \langle V_0(t'') V_0(t') \rangle
\nonumber\\ &&
+\, \langle V_0(t'') V_{\chi}(t') \rangle  
+ \langle V_0(t')V_{\chi}(t'') \rangle \big].
\label{F}
\end{eqnarray}
It is now straightforward to evaluate this function.
One  should use  the definition of the interaction potential $V_{\chi}(t)$ (\ref{Vt}), combine it with the explicit
expressions for the electron creation and annihilation operators (\ref{c_xit}) and 
apply Wick's theorem together with Eqs. (\ref{gamma_av}) to evaluate the average values of the products of these operators. 
Omitting the terms which do not depend on the counting filed $\chi$, 
one arrives at the results (\ref{FGR},\ref{Fqp_q},\ref{FJ_q}). 

\section{Dynamics of Josephson phase}

Here we briefly summarize well known results \cite{AH} on the phase dynamics in
a resistively shunted Josephson junction, which is described by Eq. (\ref{RSJ})
with $C=0$. The  probability distribution function of the phase, $\sigma(t,\varphi)$,
satisfies the Fokker-Planck equation
\begin{eqnarray}
\frac{\partial \sigma}{\partial t}+\frac{2eR_S}{\hbar}\frac{\partial}{\partial\varphi}\big[(I_x-I_C\sin\varphi)\sigma\big]
\nonumber\\
-\frac{4e^2 R_S k_B T^*}{\hbar^2} \frac{\partial^2\sigma}{\partial\varphi^2}=0.
\nonumber
\end{eqnarray}   
The stationary solution of this equation, which is periodic in phase and to which any solution approaches in the long time limit, reads
\begin{eqnarray}
\sigma(\varphi)&=&\frac{\hbar\langle V\rangle}{4\pi eR_S k_B T^*}
\int_\varphi^\infty d\varphi' e^{\hbar I_x(\varphi-\varphi') /2ek_B T^*}
\nonumber\\ &&\times\,
e^{(\cos\varphi-\cos\varphi')E_J /k_B T^*}.
\label{W0}
\end{eqnarray}
This solution is normalized as follows: $\int_{-\pi}^\pi d\varphi \sigma(\varphi)=1$.
In Eq. (\ref{W0}) we have introduced   the average voltage drop on the junction $\langle V\rangle$, which reads
\begin{eqnarray}
\langle V \rangle = 
\frac{\frac{k_B T^* \, eR_S}{\hbar}\sinh\frac{\pi\hbar I_x}{2ek_BT^*}}{\int_0^{\pi/2} d\varphi \cosh\left(\frac{\hbar I_x\varphi}{e k_BT^*}\right)I_0\left(\frac{\hbar I_C\cos\varphi}{e k_BT^*}\right)}.
\label{Vav}
\end{eqnarray}

\end{document}